\documentclass[twocolumn,preprintnumbers,amsmath,amssymb,widetext]{revtex4}
\usepackage{graphicx}
\begin{document}
\title {Coherent transport in Nb/$\delta$-doped-GaAs hybrid microstructures}
\author{F. Giazotto}
\email{giazotto@sns.it}
\author{P. Pingue}
\author{Fabio Beltram}
\affiliation{NEST-INFM \& Scuola Normale Superiore, I-56100 Pisa, Italy}

\begin{abstract}
Coherent transport  in Nb/GaAs superconductor-semiconductor microstructures is presented. The structures fabrication procedure is based on $\delta$-doped layers grown by molecular-beam-epitaxy near the GaAs surface, followed by an As cap layer to protect the active semiconductor layers during \emph {ex situ} transfer. The superconductor is then sputter deposited  \emph {in situ} after thermal desorption of the protective layer. Two types of structures in particular will be discussed, i.e.,  a \emph {reference} junction and the \emph {engineered} one that contains an additional insulating AlGaAs barrier inserted  during the growth in the semiconductor. This latter configuration may give rise to controlled interference effects and realizes the model introduced by de Gennes and Saint-James in 1963.
While both structures show  reflectionless tunneling-dominated transport, only the engineered junction shows additionally a low-temperature single marked resonance peaks superimposed to the characteristic Andreev-dominated subgap conductance.
The analysis of coherent  magnetotransport  in both microstructures is successfully performed within the random matrix theory of Andreev transport and ballistic effects are included by directly solving the Bogoliubov-de Gennes equations. The impact of junction morphology on reflectionless tunneling and the application of the employed fabrication technique to the realization of complex semiconductor-superconductor systems are furthermore discussed.  
\end{abstract}


\maketitle
\section{Introduction}
Heterostructures represent nowadays the most dynamic and rapidly growing field of semiconductor research. This success derives in particular from the great freedom they offer in the design of material energy-band diagram. With respect to  bulk structures these additional possibilities come from the addition of two basic elements: the \emph {opaque} barrier, exploited to realize quantum well and other reduced-dimensionality systems, hot-electron injectors and several heterojunction transistors, and the \emph {thin} barrier, exploited to obtain the tunnel effect (both resonant and non resonant), and superlattices. 
The present work experimentally addresses the addition a a further element, i.e., the superconducting barrier (S). This allows the introduction of an entirely new effect in the physics of the heterostructures, i.e., the \emph {Andreev reflection} \cite{andr}. This process allows the control of the nature and the phase of quasi-particles in the heterostructure: an electron injected from the normal region (N), in the present case a degenerate semiconductor (Sm), with energy lower than the superconducting gap $\Delta$ can not propagate as a single particle in the superconductor but can be reflected as a phase-matched hole in the Fermi sea of the N region, while at the same time a Cooper pair is transmitted in the superconductor. The reflected quasi-particle phase is controlled by the superconductor condensate. All this has to be contrasted to normal reflection present also in non-superconducting heterostructures. 

Thanks to Andreev reflection, dissipative currents in the normal portion of the structure are converted in dissipationless supercurrents inside the superconductor, thus allowing electric transport through the whole system.
Moreover, due to its two-particle nature, Andreev reflection is strongly affected by the NS interface transmissivity. High transparency is therefore required to observe this effect and of the associated coherent phenomena.  
The art of controlling the interface quality, in particular in the case of hybrid SmS microstructures, is still nowadays a challenging task and much effort has to be devoted to its optimization \cite{giaz1}. Achieving high interface transparency in SmS contacts, however, is hindered by the presence of interface oxide layers and Schottky barriers, and by the difference in Fermi velocities in the S and Sm layers \cite{blonder}. In order to avoid Schottky barrier formation, several authors have chosen InAs for the realization of hybrid devices. The main transparency-limiting factor in this case is interface oxidation/contamination and several methods have been explored to limit its impact \cite{kr,tak,lach}. InAs, however, is not the Sm of choice for heterostructure growth and GaAs- or InP-based systems are much more widespread and versatile. For these latter materials Schottky barriers are the dominant factor in limiting transparency.

In this paper we shall focus on some experimental results obtained  on Nb/$\delta$-doped-GaAs hybrid microstructures fabricated with a versatile and reproducible technique. GaAs is, after Si, probably the most widespread semiconductor material and at the basis of a well developed nanotechnology.  Its main peculiarities are a low effective mass value ($m^{\ast}\simeq 0.067 \,m_{e}$) that favours high bulk mobilities, and the feature that a replacement of part or all of the Ga atoms by Al atoms causes a negligible change in its lattice constant. This leads to GaAs/AlGaAs interfaces with highly reduced defect densities and in nearly perfect barriers for electron confinement. The resulting two-dimensional electron gas may yield mean free path of several tens of microns, making the GaAs/AlGaAs system the best candidate for the realization of non-superconducting nanostructures in the ballistic limit. 
These remarkable peculiarities, however, can not easily be exploited for 
 SmS applications due to the already mentioned presence of the Schottky barrier that may strongly hinder the manifestation of Andreev reflection. Nevertheless, GaAs can still be employed with superconductors by heavily doping the semiconductor surface layers in order to achieve high interface transmissivity (this technique is commonly referred to  as $\delta$-doping technique) \cite{kast,tab}. This is indeed a practical and effective way to use GaAs in the contest of SmS structures avoiding annealed superconducting contacts \cite{gao,mar}, and will be demonstrated in the following. 
More in particular, we shall examine two Nb/$\delta$-doped-GaAs microstructures, that will be referred to as the SN \emph {reference} \cite{giaz2}  and the SNIN \emph {engineered} \cite{giaz3} structure.
The latter  differs from the reference junction for the additional presence of an Al$_{0.3}$Ga$_{0.7}$As barrier inserted during the growth process. 

The paper is organized as follows. 
In Section II we shall describe the structures characteristics  and some details of the employed fabrication technique. The electrical behavior of the SN reference junctions will be addressed in Section III, where we will demonstrate the observation of reflectionless tunneling(RT)-dominated transport. Such results will be discussed within a random matrix approach. We shall report here the effectiveness of the employed fabrication protocol for the implementation of more complicated systems, i.e., the SNIN engineered junctions. In Section IV we shall present the experimental evidence in these latter structures of resonant states  induced by the presence of the AlGaAs insulating barrier. The observed behavior is of the de Gennes-Saint-James type and reveals how it is possible to tailor these semiconductor heterostructures in order to observe coherent transport effects. The electrical behavior of these junctions will be described within a random matrix formalism and ballistic effects will be included directly solving the Bogoliubov-de Gennes equations in a model potential profile. Our description of the system is confirmed by the observed temperature and magnetic-field dependence.
The conclusions and some additional remarks will be given in Section V.

\section{Samples fabrication and experimental details}
\begin{figure}[h!]
\begin{center}
\includegraphics[width=7.0cm]{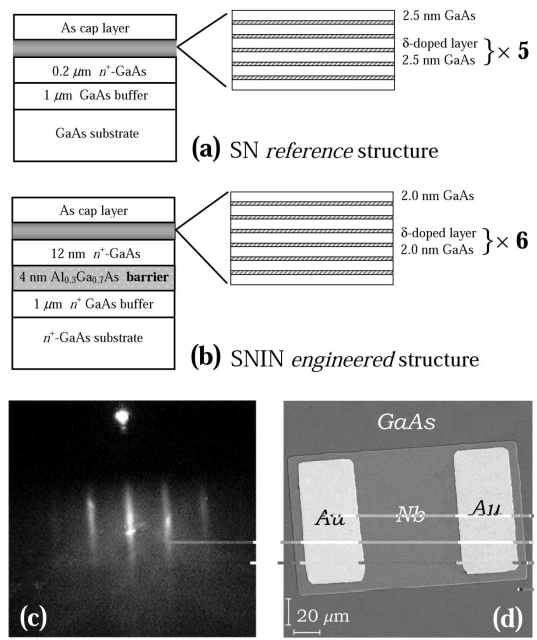}
\end{center}
\caption{Schematic cross section of the semiconductor heterostructures studied in this work: (a) the SN \emph {reference} and (b) the SNIN \emph {engineered} structure. (c) RHEED pattern image of one of the decapped samples just before the deposition of Nb. This pattern was recorded at 10 KeV and at a current of 1.4 A. (d) Scanning electron micrograph of one of the measured devices 
}
\label{ARscheme}
\end{figure}
For the semiconducting portion of the SN \emph {reference} structure (see Fig. 1(a)), a 200-nm thick n-GaAs(001) epilayer (nominal Si doping n\,=\,4.7\,$\times 10^{18}$\,cm$^{-3}$) was initially grown by molecular beam epitaxy (MBE) on an undoped GaAs buffer and a semi-insulating GaAs(001) wafer. This was followed by the growth of 15 nm of GaAs doped by a sequence of five $\delta$ layers (nominal Si concentration $3 \times 10^{13}$ cm$^{-2}$ per layer) spaced by 2.5 nm.
The semiconductor portion of the SNIN {\it engineered} structure (see Fig. 1(b)), instead, consists of a 1 $\mu$m thick $n$-GaAs(001) buffer layer Si-doped nominally at $n= 2\times 10^{18}$ cm$^{-3}$ grown by  MBE on a $n$-GaAs(001) substrate, followed by a 4 nm thick Al$_{0.3}$Ga$_{0.7}$As barrier. This was followed by the growth of a 12 nm thick GaAs(001) epilayer Si-doped at $n= 2\times 10^{18}$ cm$^{-3}$ and by 14 nm of GaAs doped by a sequence of six Si $\delta$-doped layers spaced by 2 nm.
The thickness of the GaAs epilayer sandwiched between the superconductor and the AlGaAs barrier was selected in order to have an experimentally-accessible single quasi-bound state below the superconductive gap, and the Si $\delta$-doped layers at the Nb/GaAs  interface were employed to achieve the required transmissivity.
A similar doping scheme was successfully used by Taboryski and co-workers to lower the contact resistance in fully-MBE-grown Al/GaAs(001) junctions \cite{tab}.
The choice of \emph {in-situ} junction formation, however, severely limits the superconductor materials usable. In our case, on the contrary, following the growth of the top $\delta$-doped GaAs layer, a 1-$\mu$m-thick amorphous As cap layer was deposited at -20$^\circ$C on both structures to protect the surface during transfer in air to an ultra-high-vacuum  (UHV) sputter-deposition/surface analysis system (the typical base pressure of the UHV fabrication system is about  $1.2\times 10^{-10}$ Torr). The samples where then heated at about 400$^\circ$C to desorb the cap layer in order to achieve a sharp GaAs(001) 2\,$\times$\,4 reflection-high-energy-electron-diffraction (RHEED) pattern.
An example of the corresponding RHEED pattern image obtained on the SN reference structure is shown in Fig. 1(c). The RHEED pattern of the SNIN engineered junction appeared quite similar.
Nb overlayers 100-nm-thick were fabricated \emph {in situ} by dc-magnetron sputter-deposition at a deposition rate of 3.5 nm/s. Moreover, substrate temperature during Nb deposition was kept at $\approx$ 200\,$^\circ$C to promote film adhesion. Typical 100-nm-thick Nb films displayed a transition temperature $T_{c}$ of 9.28\,K and a residual resistivity ratio at 10 K (RRR$_{10}$) as high as 60 (data not shown).

Rectangular 100$\times$160\,$\mu$m$^2$ Nb/GaAs junctions were patterned by 
standard photolithographic techniques and reactive ion etching (RIE) using a 
CF$_{4}$+O$_{2}$ gas mixture. Two additional 
90$\times$45\,$\mu$m$^2$ Ti/Au bonding pads were e-beam evaporated on top of 
each Nb contact to allow 4-wire measurements (Figure 1(d) shows the scanning electron micrograph of one of the measured devices). 
In the case of  SN reference junctions, the electrical characterizations were performed with two leads on the Nb electrode under interest and the other two connected to two separate neighboring contacts located symmetrically with respect to the junction considered. In this way our conductivity data reflect only the junction properties with no influence from the series resistance of the semiconductor film.
We have also experimentally determined the influence of the leakage paths around the contacts by removing the semiconductor layer above and below the contact strip (``contact-end test structure'' \cite{sch}). The correction to the contact resistance appeared to be negligible on the scale relevant to the experiment. 
For what concerns  SNIN devices, differently,  4-wire measurements were performed with two leads on the junction under study and the other two connected to the sample back contact, thus allowing to probe the vertical transport characteristic intrinsic to the structure. 

Magnetoconductance measurements
of the junctions were performed in a closed-cycle $^3$He cryostat 
equipped with a superconducting magnet from 0.3 K to temperatures larger than $T_{c}$, and current-driven  measurements were performed employing a high-resolution current source. Furthermore, in order to preserve any sharp features in the junction characteristics, special care was taken in  data acquisition using low measurements speed and small current stepwidths.

\section{ SN  reference junctions: reflectionless tunneling-dominated transport}
\begin{figure}[h!]
\begin{center}
\includegraphics[width=4.8cm,angle=-90]{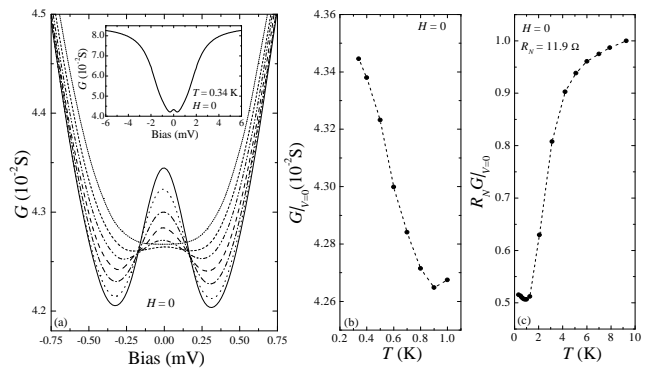}
\end{center}
\caption{(a) Differential conductance vs voltage at several temperatures and no applied magnetic field. From bottom to top, the conductance spectra were recorded at 0.34, 0.50, 0.60, 0.70, 0.80, 0.90, 1.00 K. The inset shows the conductance at 0.34 K in a wider bias range. The reflectionless tunneling enhancement of the conductivity appears as a peak around zero bias. (b) Temperature dependence of the zero-bias differential conductance  that shows the extreme sensitivity to temperature of the reflectionless tunneling peak. (c)Temperature dependence of the  normalized zero-bias differential conductance  up to Nb critical temperature. In (b) and (c) dashed lines are guides for the eye. 
}
\label{ARscheme}
\end{figure}
In order to 
determine carrier concentration and Hall mobility of the SN reference structure a portion of the 
semiconductor was not Nb coated and, following As desorption, it was patterned into Hall bars by removing the entire epilayer. At 0.34\,K we obtained $n=3.1\times 10^{18}$\,cm$^{-3}$ 
and $\mu = 1.6\times 10^{3}$\,cm$^{2}$/V s. From these values it was possible 
to estimate electron mean free path (${\ell}\,\approx 50$\,nm) and  
thermal coherence length at $T$\,=\,0.34\,K, 
$\xi _{T} (T)=\sqrt{\hbar D /2\pi k_{B} T}\approx\,0.21\,\mu$m, 
where $D=1.2\times 10^{2}$\,cm$^{2}$/s is the three-dimensional diffusion constant. Additionally, the single-particle phase coherence length, 
$\ell_{\phi}\leq 1$\,$\mu$m, was also estimated from weak localization 
magnetoresistance measurements, following Ref. \cite{kut}.

Figure 2(a) shows a typical set of differential-conductance $vs$ bias 
characteristics ($G(V)$) of one of the junctions measured at various 
temperatures in the 0.34\,-\,1.00\,K temperature range. The symmetry of the
characteristics together with the large value of the conductivity at
low bias at the lowest temperatures are a clear indication of the 
transparency of the Nb/GaAs junctions fabricated (see also the inset of Fig. 2(a)
where the conductance spectrum at 0.34\,K up to biases well above the superconducting energy gap  is plotted). The fabrication procedure adopted yielded an excellent contact-to-contact uniformity leading to $G(V)$ characteristics reproducible within few \% in all the junctions measured. The data display the typical behavior of a diffusive non-tunnel SmS junction. In particular, at the lowest temperatures measured, a marked peak is observed in the differential conductance spectrum at zero bias. As we shall argue, this is the direct manifestation of reflectionless tunneling (RT) \cite{kast,marmor,ben1};
RT is indeed known to be due to constructive interference of quasiparticles coherently backscattered withing the diffusive region towards the SmS interface barrier \cite{wees}. This effect gives rise to a sizable enhancement of conductivity around zero-bias, and its coherent nature makes it very sensitive to electric and magnetic fields and to the temperature.
Furthermore,  it will be very useful for the analysis of the junction properties. 
\begin{figure}[h!]
\begin{center}
\includegraphics[width=5.3cm,angle=-90]{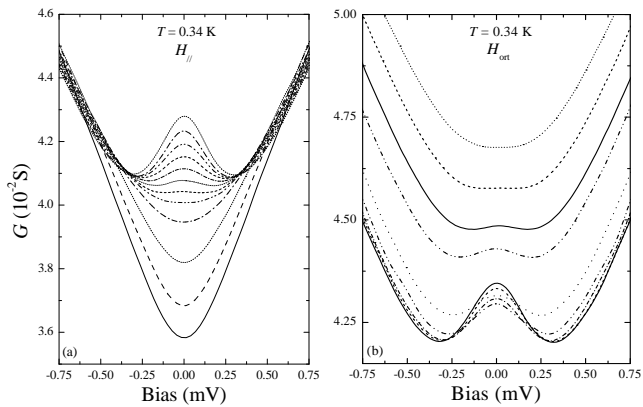}
\end{center}
\caption{Differential conductance vs voltage at 0.34 K for various magnetic field values. (a) Magnetic field applied in the plane of the junctions; from top to bottom, data were recorded at 0, 2, 3, 4, 5, 6, 7, 8, 10, 16, 30, 100 mT. (b) Magnetic field applied perpendicularly to the junction plane; from bottom to top, data were recorded at 0, 1, 2, 3, 4, 5, 6, 7, 8 mT. In this case, the conductance behavior is to be ascribed to the appearance of fluxons in the mixed state of Nb. 
}
\label{ARscheme}
\end{figure}
In fact it is not straightforward to obtain reliable junction parameters by studying the Andreev pattern in diffusive junctions \cite{giaz4,neu}. On the contrary, when one such zero-bias conductance peak (ZBCP) is observed in these systems, much information can be obtained by
examining its intensity and its temperature and magnetic-field 
dependence \cite{marmor,ben1}. The temperature dependence of the RT zero-bias enhancement is shown for clarity in Fig. 2(b), while Figure 2(c) shows the normalized  zero-bias conductance dependence ($R_N =11.9 \,\,\Omega$ is the junction normal state resistance) in the whole temperature range up to Nb critical temperature. With the exception of the low temperature range,  where the RT enhancement persists, this behavior is that one expected for a  junction in the diffusive regime.

The electrical properties of our junctions can be analyzed within the random matrix theory of phase-coherent Andreev reflection proposed by Beenakker \cite{ben2}.
He studied a model-junction consisting of a normal and a superconductor reservoirs linked by a disordered normal-metal region of length $L$ and width $W$. 
Between this disordered region and the superconductor reservoir the model-system includes a potential barrier characterized by a mode-independent transmission probability ($\Gamma$). In this system, under appropriate conditions an enhanced conductivity around zero bias was predicted to occur with respect to the {\it classical} Andreev reflection behavior leading to a ZBCP. The width of the latter (i.e. the bias voltage at which the RT correction to the conductivity vanishes) is given by 
$V_{c}=(\pi /2)\hbar v_{F} \ell/eL^{2}$, where $v_{F}$ is the Fermi velocity in the normal region. Furthermore, upon application of a magnetic field ZBCP is suppressed for field intensities larger than $B_{c}=h/eLW$. In real systems, at finite temperatures, $L$ and $W$ are to be replaced by $\xi _{T}(T)$, if it is smaller \cite{marmor,ben1}.  
In order to analyze our data within this model, we can first of all determine the experimental value of $V_c$ by examining the data in Fig. 3(a) where the evolution of the ZBCP is well displayed thanks to the limited influence of the in-plane magnetic field on the classical conductivity. The RT correction persists up to about $V_c\approx 0.6$\,mV corresponding to a characteristic length $L \approx$\,0.25\,$\mu$m. As predicted in Refs. \cite{marmor,ben1} , this value is in good agreement with the estimated thermal coherence length at the same temperature.

Further insight on the interference effects leading to RT can be gained
observing the magnetic-field dependence of the ZBCP.
Figure 3 shows a set of $G(V)$'s at $T=0.34$\,K for several 
magnetic fields applied (a) in the plane of and (b) perpendicularly to the plane of the  junction. At relatively weak magnetic fields $\sim$\,mT the peak is suppressed and the subgap conductance dip shrinks due to the magnetic-field-induced suppression of the superconductor energy gap $\Delta$. At higher values of the applied magnetic field the overall subgap conductance increases approaching its normal-state value (data are not shown for clarity). From the in-plane-field data, we can determine with good accuracy the experimental critical magnetic field $B_{c} \approx 100$\,mT corresponding to RT suppression and to the minimum measured value of the zero-bias conductivity. 
This value is in good agreement with the expected theoretical value $B_{c} \approx 80$\,mT obtained from the estimated thermal coherence length at $T=0.34$ K, $\xi _{T}\approx 0.21\,\mu$m, confirming the physical origin of the observed conductance behavior.

It is interesting to compare the conductance behavior caused by different field orientations. For perpendicular fields the overall subgap conductance increases at much smaller fields as compared to the in-plane configuration. We attribute this effect to the type-II nature of Nb and  to the consequent appearance of vortices in the mixed state. The presence of these normal regions within the junction leads to increased subgap conductance  due to the lower resistance of the normal vortex region with respect to the superconducting portions. A similar behavior was also observed by Quirion and co-workers \cite{quir} in planar TiN-Si junctions.
A quantitative estimate of the normal-zone contribution to the conductivity is however hindered by the uncertainty about the effective junction area contributing to the transport. We should like to stress the importance of  
careful analysis of the in-plane field curves for a quantitative determination of the RT correction to zero-bias classical conductivity without any empirical extrapolation. 
\begin{figure}[h!]
\begin{center}
\includegraphics[width=5.2cm,angle=-90]{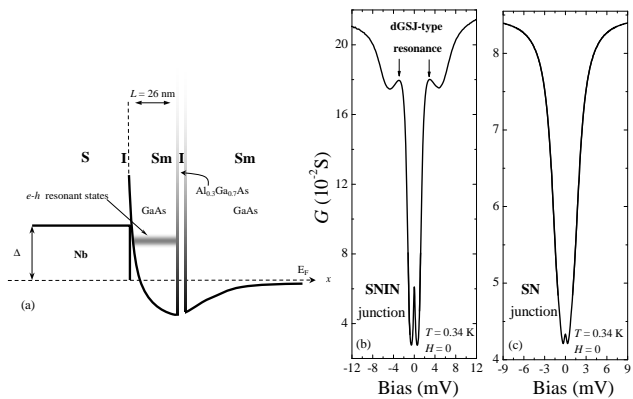}
\end{center}
\caption{(a) Sketch of the energy-band diagram of the Nb/GaAs/AlGaAs/GaAs SNIN engineered structure. The shaded area represents a de Gennes-Saint-James quasi-bound state confined between the superconductor and the AlGaAs barrier. Differential conductance vs voltage at 0.34 K for the engineered SNIN junction, (b), and for the reference SN junction, (c). Both curves show reflectionless tunneling enhancement around zero bias, while only in (b) (marked by arrows) a finite-bias, subgap de Gennes-Saint-James peak is present. 
}
\label{ARscheme}
\end{figure}
The quantitative analysis of the ZBCP intensity, within the model adopted so far can yield an estimate of the junction transmission probability ($\Gamma$). However, the transmissivity calculated from the ratio between the resistance values taking into account the phase-coherent AR and the classical behavior for SN systems are incompatible with the phenomenology observed. In fact, for our junction properties ($L >> \ell$), the observed phenomenology (i.e. $increased$ conductance at zero bias) is expected for $\Gamma \leq 0.4$ while the above procedure leads to significantly higher values. This deviation in the observed intensity of the ZBCP must be linked to a significant suppression of the RT-induced enhancement of the conductivity. We believe that this effect stems from junction inhomogeneities. Several authors have reported that only a very small fraction (10$^{-2}$--10$^{-4}$) of the junction physical area determines the transport properties of large diffusive contacts \cite{giaz1,kast,gao,huff}. Additionally, the very good contact-to-contact uniformity we observed indicates that the lateral scale of the higher-transparency regions must occur on an even smaller characteristic length. This peculiar junction morphology leads to a drastic reduction in the number of the reflectionless paths available for ZBCP build-up and to the observed suppression of the ZBCP intensity.

\section{SNIN engineered junctions: de Gennes-Saint-James resonant transport}

In Section III we demonstrated the impact  of scattering centers in the normal region on electron transport. These can indeed significantly affect junction properties leading  to marked coherent-transport phenomena such as the observed RT in  SN reference junctions. 
\begin{figure}[h!]
\begin{center}
\includegraphics[width=5.5cm,angle=-90]{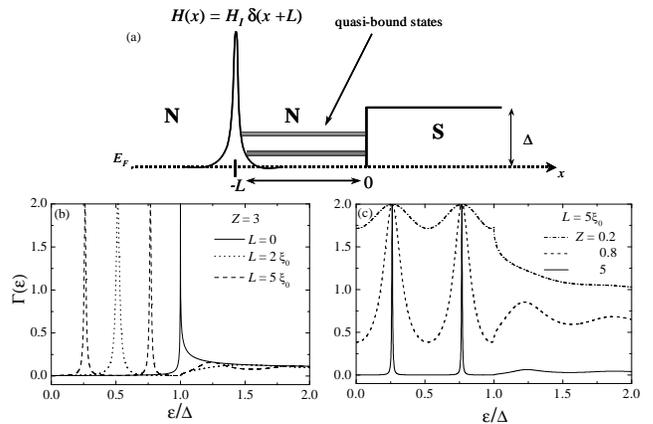}
\end{center}
\caption{(a) Energy-band diagram of the one-dimensional SNIN ballistic model employed to analyze the experimetal data (see text). (b) Calculated total transmission coefficient vs energy at zero temperature for fixed barrier strength and some N interlayer thicknesses. More subgap quasi-bound states are allowed incresing such thickness. (c) Calculated total transmission coefficient vs energy at zero temperature for fixed N interlayer thickness and various barrier strength values. Increasing barrier strength yields to longer resonant states lifetimes. 
}
\label{ARscheme}
\end{figure}
One particularly interesting case is that of a single scatterer represented by an insulating barrier (I) inserted in the structure during growth in the normal region. This configuration can give rise to controlled interference effects. Among these, one of the most intriguing is represented by de Gennes-Saint-James (dGSJ)  resonances \cite{dgsj} in SNIN systems where the N interlayer is characterized by a constant pair potential. The case of a null pair potential is especially relevant to the present case. 
Multiple reflections off the superconductive gap (i.e. Andreev reflections) and off the insulating barrier (i.e. normal reflections) may give rise  to quasi-bound states \cite{dgsj,rb} that manifest themselves as conductance resonances. Transport resonances linked to similar multilayer configuration were observed experimentally in \emph {all-metal} structures \cite{row,wong,tess}, thus providing elegant evidence of quasi-particle coherent dynamics in SN systems. A qualitative sketch of the energy-band diagram of our SNIN \emph {engineered} structure is depicted in Fig. 4(a).

In Fig. 4 are shown the measured differential conductance $vs$ bias ($G(V)$) for the resonant structure (SNIN engineered junction, (b)) and for the reference junction (SN structure, (c)) at $T=0.34$ K. Comparison of the two characteristics clearly shows the presence of a marked subgap conductance peak in the SNIN, resonant device. The resonance is superimposed to the typical Andreev-dominated subgap conductance. The symmetry of conductance and the  ZBCP peculiar to  RT further demonstrate the effectiveness of the employed fabrication protocol.
\begin{figure}[h!]
\begin{center}
\includegraphics[width=5.2cm,angle=-90]{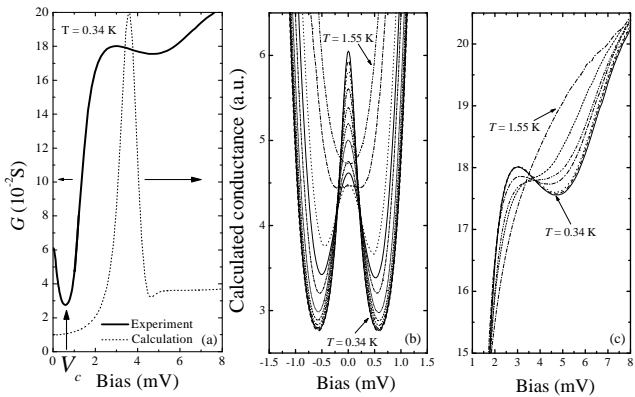}
\end{center}
\caption{(a) Comparison at 0.34 K between the experimental differential conductance (solid line) and the numerical simulation from Eq.1 (dotted line). Differential conductance vs voltage at several temperatures in the 0.34 to 1.55 K range  for the zero-bias conductance peak, (b), and for the de Gennes-Saint-James resonance peak, (c). Data were taken at the following temperatures: (b), from bottom to top, 0.34, 0.45, 0.55, 0.65, 0.75, 0.85, 0.95, 1.05, 1.26, 1.45, 1.50, 1.55 K; (c), from bottom to top, 0.34, 0.55, 0.65, 0.85, 1.05, 1.26, 1.55 K. 
}
\label{ARscheme}
\end{figure}
Quantitative determination of the resonant transport properties in such a system is not trivial as it can be inferred by inspecting Fig. 4(a) and considering the diffusive nature of the normal regions. dGSJ-enhancement, however, is an intrinsically ballistic phenomenon so that its essential features can be captured with relative ease. One study particularly relevant for this system was performed in Ref. \cite{rb}. In the context of ballistic transport a one-dimensional SNIN structure (see Fig. 5(a)) was studied as a function of the N interlayer thickness $L$ and it was demonstrated that resonances (i.e., quasi-bound states) can occur in the subgap conductance spectrum for suitable geometric conditions. The insulating barrier  was simulated by a $\delta$-like potential $H(x)= H{_I} \delta (x+L)$.  Customarily the barrier strength is described by the dimensionless coefficient $Z= H{_I}/ \hbar v_F$, where $v_F$ is the electron Fermi velocity \cite{btk}, and in the following we shall make use of it to characterize our system.
The key-results of the analysis are: ({\emph i}) the number of resonances increases for larger thickness $L$ of the metallic interlayer and is virtually independent of $Z$; (\emph {ii}) the energy-width of such resonances decreases for increasing $Z$. All this is better highlighted in Fig. 5(b) and 5(c). Figure 5(b)  shows the calculated total transmission coefficient $\Gamma(\epsilon)=[1+A(\epsilon)-B(\epsilon)]$, where $A(\epsilon)$ and $B(\epsilon)$ are the energy-dependent Andreev  and normal reflection transmission probabilities for the whole SNIN structure obtained through the solution of the Bogoliubov-de Gennes equations, at $T=0$ for $Z=3$ and for some values of the N interlayer thickness $L$ expressed in units of the superconducting coherence length, $\xi _0 =\hbar v_F /2 \Delta$. In these simulations we used, as an example, parameters typical for metals, i.e.,   the Fermi energy $\mu =1$ meV, the quasi-particle mass equal to the free electron mass $m_e$ and $\Delta = 10$ meV. From this choice follows that $\xi _0 =196$ {\AA}. The subgap resonances reach a transmission value $\Gamma =2$ due to resonant Andreev reflection at the SN interface. Increasing the N interlayer thickness allows the formation of more quasi-bound states, like in  ordinary normal resonant tunneling. When the barrier is located at the SN interface we recover the expected Blonder-Tinkham-Klapwijk result \cite{btk} for an opaque junction.
In Fig. 5(c) is shown the calculated $\Gamma(\epsilon)$ at $T=0$ for some values of barrier strength $Z$ and for fixed $L=5\,\xi _0$. The other parameters of the simulations are the same as in Fig. 5(b). Increasing $Z$ makes the resonances much more sharp: indeed it is possible to demonstrate \cite{rb} that the energy width of the resonance, $\delta \epsilon$, is of the order of $\delta \epsilon \simeq \hbar v_F/2L(1+Z^2 )$; therefore, incresing the barrier strength, decreases the resonance energy width up to a {\it monochromatic} or fully-bound state in the limit of infinite barrier.  
The one-dimensional differential conductance $G(V)$ at temperature $T$ can be expressed as \cite{btk}
\begin{equation}
G(V) =G_{0} \int_{-\infty}^{\infty}\Gamma (\epsilon)\,f_{0}^{2}(\epsilon -qV) \, e^{\frac{\epsilon-qV}{k_{B}T}}d \epsilon\,\,,
\end{equation}
where $G_{0}=q^2/\pi k_{B}T\hbar$, $qV$ is the incoming particle energy and $f_{0}(\epsilon)$ is the equilibrium Fermi distribution function.  
This model is rather idealized, but is a useful tool to grasp the essential features of our system such as number and position of resonances. We confined our calculation to the one-dimensional case in light of the results by S. Chaudhuri and P. F. Bagwell \cite{cb}, that showed the insensitivity to dimensionality of the essential properties of transport resonances. These are determined by the  inspection of the $G(V)$ behavior. 
In order to apply it we must first determine  parameters such as barrier strength, electron density and mean free path. Also, any quantitative comparison with experiment requires us to estimate the sample series resistance, which influences the experimental energy position of the resonance peak. The above parameters can be obtained from an analysis of the RT-driven ZBCP and from Hall measurements.

We performed Hall measurements at 1.5 K and obtained carrier density $n\simeq 4\times 10^{18}$\,cm$^{-3}$ and mobility $\mu \simeq 1.5\times 10^{3}$\,cm$^{2}$/V s. These data allow us to estimate the thermal coherence length $\xi _T(T) =\sqrt{\hbar D/2\pi k_{B}T}=\,0.13\,\mu$m$/\sqrt{T}$, where $D=1.34\times 10^2$\,cm$^2$/s is the diffusion constant, and the electron mean free path ${\ell}\,\approx 48$\,nm.  $ {\ell}$ compares favorably with the geometrical constraints of the structure (as a matter of fact $L=26$ nm $< {\ell}$) and further supports our ballistic analysis.
The ZBCP can be described following the analysis of Refs. \cite{marmor,ben1} and described in Section III. 
At 0.34 K, $\xi_{T} \approx 0.22$\,$\mu$m and in our junctions we calculate $V_c \simeq 200$\,$\mu$V. Comparison with the experimental value $V_c^{exp} \simeq 600$\,$\mu$V in Fig. 6(a) (see solid line), allowed us to estimate the series resistance contribution to the measured conductivity. This rather large effect stems mainly from the AlGaAs barrier and the sample back-contact resistance.

\begin{figure}[h!]
\begin{center}
\includegraphics[width=5.5cm,angle=-90]{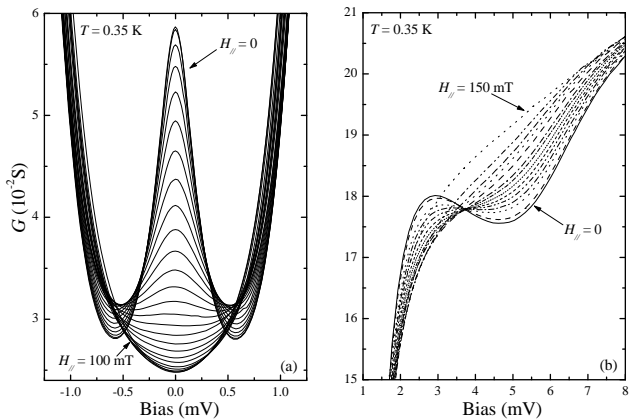}
\end{center}
\caption{Differential conductance vs voltage at 0.35 K for several values of the magnetic field applied in the plane of the junction. Dependence (a) of the zero-bias conductance peak, and (b) of the de Gennes-Saint-James resonance peak. In (a) data were taken in the 0 to 100 mT range with a 4 mT step, while in (b) at 0, 10, 20, 25, 30, 35, 40, 45, 50, 60, 70, 80, 100, 150 mT. 
}
\label{ARscheme}
\end{figure}
Following Eq. 1 we calculated the conductance for the nominal N-thickness value ($L = 26$ nm) and $Z = 1.4$, as appropriate for the AlGaAs barrier. It is indeed straightforward to determine $Z\simeq 1.4$, taking into account the Al$_{0.3}$Ga$_{0.7}$As/GaAs discontinuity, Fermi energy and barrier thickness. We emphasize, however, that the essential features such as the number and energy position of the dGSJ resonances are virtually independent of Z and are controlled instead by the value of $L$. Our calculations yield a single conductance peak at energies corresponding to about $0.8\,\Delta$. By including the above-determined series resistance contribution, the resonance peak is positioned at about $3.5\,$mV (see Fig. 6(a), dotted line, calculated at $T=0.34$ K). 
The corresponding experimental value is about 3 mV (solid line in Fig. 6(a)), but the observed energy difference is well within the uncertainty resulting from the determination of the series resistance.
The results of our model calculations strongly support our interpretation of the experimental structure in terms of dGSJ resonant transport.
Previous results on all-metal films generally showed more complex structures resulting from several dGSJ peaks, consistent  with the wider N-regions employed and the larger Fermi wave-vector  $k_F$ values characteristic of metallic systems. Our calculations show that for our material system configurations presenting more than one peak (i.e., more quasi-bound states) are not experimentally accessible. 
Indeed,  they show  that a second peak occurs for N-region thickness exceeding 120 nm. This length unfavorably compares with the available quasi-particle coherence length $\xi_{T}$  particularly in light of the known four-traversal requirement for dGSJ quasi-bound state formation \cite{wolf}.

Ordinary resonant tunneling in the normal double-barrier potential
schematically shown in Fig. 4(a) cannot explain the observed subgap structure. This is indicated
by the symmetry in the experimental data for positive and negative bias and is further proven
by the temperature and magnetic field dependence of the differential conductance. 
Figure 6 shows a set of $G(V)$s measured in the 0.34-1.55 K range for the ZBCP, (b), and for the resonance peak, (c). Both effects show a strong dependence on temperature and at $T=1.55$ K are totally suppressed. This temperature value is within the range where RT suppression is expected \cite{kast,giaz2,post,sanq}.
At higher temperatures, the conductance in the resonance region resembles that one of the reference SN junction of Fig. 4(c).
Notably, the ZBCP and the resonance peak disappear at the same temperature, hence indicating the coherent nature of the observed effect.

\begin{figure}[h!]
\begin{center}
\includegraphics[width=5.5cm,angle=-90]{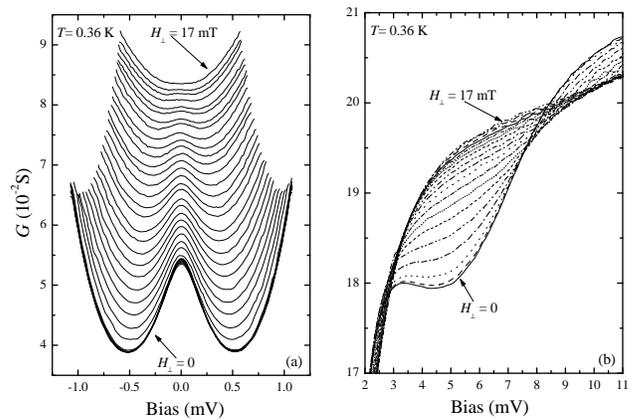}
\end{center}
\caption{Differential conductance vs voltage at 0.36 K for several values of the magnetic field in the 0 to 17 mT range applied perpendicularly to the junction plane. Dependence (a) of the zero-bias conductance peak, and (b) of the de Gennes-Saint-James resonance peak.  In (a) data were recorded with a 0.5 mT step, while in (b) with a  1 mT step. 
}
\label{ARscheme}
\end{figure}
Further confirmation of the nature of the resonance peak can be gained observing its dependence from the magnetic field. Figure 7 shows $G(V)$ at $T=0.35$\,K for several values of the magnetic field applied in the plane of the junction for the ZBCP, (a), and for the resonance peak, (b). 
The measurements confirm the known sensitivity of ZBCP to the magnetic field \cite{marmor,ben1}, and clearly indicate that the dGSJ resonance is easily suppressed for critical fields of the order of 100 mT. 
We also investigated the perpendicular field configuration (see Fig. 8) for the ZBCP, (a), and for the resonance peak, (b), and also in this case 
 the resonance and the ZBCP displayed a similar dependence. 
Such behavior is fully consistent with dGSJ-related origin but is not compatible with a normal resonant tunneling description of the data \cite{capasso}. Moreover, in the perpendicular configuration, the differential conductance spectra show a behavior similar to that one reported for the SN reference junctions in Fig. 3(b), i.e., the appearance of fluxons in the mixed state of Nb.

\section{Concluding remarks}

In conclusion, we have reported on reflectionless tunneling-dominated transport in Nb contacts on GaAs (i.e., in SN reference junctions) and presented an effective and reproducible fabrication technique. The observed regime was studied as a function of temperature and magnetic field. Data were analyzed within the random-matrix theory of phase-coherent Andreev reflection and the impact of junction nonuniformity on reflectionless tunneling was underlined.
The fabrication technique presented is compatible with the most widespread MBE systems and can be easily exploited for the implementation of novel hybrid SmS mesoscopic systems by tailoring the band-gap profile of the AlGaAs/GaAs heterostructures on which the Nb film is deposited and patterned. 
This was indeed developed in the SNIN engineered junctions, where we  experimentally observed de Gennes-Saint-James resonant states in Nb/GaAs/AlGaAs/GaAs hybrid microstructures. 
In these junctions the electric transport was studied as a function of temperature and magnetic field and was successfully described within the ballistic model of Riedel and Bagwell \cite{rb}.
To the best of our knowledge this result represents the first demonstration of  de Gennes-Saint-James resonant transport in SmS hybrid structures and was made possible by the fabrication procedure adopted in this work.
The present results confirms that the Nb/GaAs/AlGaAs system is a good candidate for the implementation of complex mesoscopic structures that can take advantage of the mature AlGaAs nanofabrication technology. Such structures may represent ideal prototype systems for the study of coherent transport and the implementation of novel hybrid devices.

The experimental results reported in this paper  highlight some interesting issues.
The first one is related to the technique that we have adopted and refers in particular to the high reproducibility of the structures and to their large homogeneity on the lateral scale. Indeed, both in Nb/GaAs and Nb/GaAs/AlGaAs/GaAs structures the behavior was consistently the same for all the measured junctions. This is intrinsic, we believe, to the Sm surface treatment before the contact with the superconductor.
The accurate lattice reconstruction achieved after the substrate annealing yields highly controlled  surface conditions, avoiding those problems encountered in treating   semiconductors with other  methods, i.e., typically with wet or dry etching. Furthermore we would like to stress that this is a fairly simple technique. Even the choice of metals to be deposited  on the semiconductor substrate is in principle extremely wide and allows to test several superconductor combinations. The drawback of this method resides in the fact that it can not be employed in those situation where the semiconducting active layer is underneath the surface. The exploitation of high-mobility semiconductor heterostructures  is thus precluded.

Another important point  is related to the GaAs system that proved to be a good choice for the realization of hybrid devices. Junctions showed good interface transmissivity of the order of $\Gamma=0.2 \div 0.4$ and in principle it should be possible to engineer the GaAs doping profile in order to further enhance the interface transparency. This represents  an interesting issue in order to establish an  optimized protocol for interface transparency in  hybrid structures fabricated with such material.
The combination with AlGaAs  revealed the effectiveness of tailoring the structure band profile in order to explore phase-coherent effects. The   flexibility offered by the GaAs/AlGaAs system seems  indeed ideally suited  to design more elaborate heterostructures to probe the effect of the superconducting state.
In light of these results it seems natural to reckon the possibility of extending this technique to MBE-grown InAs and In$_{0.77}$Ga$_{0.23}$As layers, where the lack of the Schottky barrier together with low values of the effective mass as compared to GaAs could in principle allow more easily the implementation of hybrid structures in the ballistic transport regime.

\section{Acknowledgment}
The authors acknowledge the financial support of INFM under the PAIS project EISS (Eterostrutture Ibride Semiconduttore-Superconduttore).



\end{document}